\documentclass[onecolumn,preprintnumbers,eqsecnum,12pt]{revtex4}
\usepackage{amsfonts}
\usepackage{amssymb}
\usepackage{amsmath}
\usepackage{epsf}
\usepackage{graphicx}
\usepackage{dcolumn}
\usepackage{bm}

\begin{document}

\title{Tension term, interchange symmetry, and the analogy of energy and
tension laws of the AdS soliton solution}

\author{Ya-Peng Hu$^{a,b,}$\footnote{e-mail address:
yapenghu@itp.ac.cn}}

\address{$^{a}$
Key Laboratory of Frontiers in Theoretical Physics, Institute of
Theoretical Physics, Chinese Academy of Sciences, P.O. Box 2735,
Beijing 100190, China}
\address{$^{b}$
Graduate School of the Chinese Academy of Sciences, Beijing 100039,
China}

\vspace*{1.cm}
\begin{abstract}
In this paper, we reconsider the energy and tension laws of the
Ricci flat black hole by taking the contribution of the tension term
into account. After this considering and inspired by the interchange
symmetry between the Ricci flat black hole and the AdS soliton
solution which arises from the double analytic continuation of the
time and compact spatial direction, we find out the analogy of the
energy and tension laws of the AdS soliton solution. Moreover, we
also investigate the energy and tension laws of the boosted Ricci
flat black hole, and discuss the boosted AdS soliton solution.
However, although there is the same interchange symmetry between the
boosted Ricci flat black hole and boosted AdS soliton, the analogy
of laws of the boosted AdS soliton solution may be of no sense for
the existence of the closed timelike curves and conical singularity.
In spite of that, the conserved charges such as the energy and
momentum of the boosted AdS soliton are well-defined, and an
interesting result is that its energy is lower than that of the
static AdS soliton. On the other hand, note that although the laws
obtained above are the same as those of the asymptotically flat
case, the underlying deduced contents are different. Thus, our
results could also be considered as a simple generalization to the
asymptotically AdS case. Moreover, during the calculation, we find
that there may be a new way to define the gravitational tension
which can come from the quasi-local stress tensor of the
counter-term method.
\end{abstract}

\maketitle

\newpage

\section{Introduction}

It is well-known that the positive energy theorems ensure the energies of
the solutions approaching AdS spacetime globally cannot be negative\cite{Yau,Witten,Gibbons}%
. However, if the considering spacetimes are locally asymptotically
AdS but not globally, the positive energy theorem may not hold. The
Horowitz-Myers AdS soliton solution is this kind of particular
solution~\cite{Horowitz}. This AdS soliton solution is important not
only for its negative energy, but also for the agreement with the
Casimir energy in the field theory viewed from the AdS/CFT
correspondence\cite{Maldacena}. Furthermore, it has also been found
that there is a similar phase transition like the Hawking-Page phase
transition between the Ricci flat black hole and the AdS soliton
solution, and it could be connected with the
confinement/deconfinement phase transition in QCD
\cite{Hawking,Witt,Cai}.

Although many properties of this AdS soliton solution have been
studied, the analogy of its energy and tension laws is absent, and
it is simply because its entropy is zero and the period of the
imaginary time is arbitrary. Recently, inspired from the interchange
symmetry between the KK bubble and the corresponding black hole
which are all asymptotically flat, D.Kastor et al obtained some
interesting results of the KK bubble after defining some new
quantities such as its surface gravity and the area of the KK
bubble~\cite{Kastor}(Note that the surface gravity here is
associated with the spacelike Killing field which translates around
the compact spatial coordinate, and more details can be found in
\cite{Gibbons2}). In our paper, viewed from the similar interchange
symmetry between the AdS soliton solution and the Ricci flat black
hole, we first reinvestigate the energy and tension laws of the
Ricci flat black hole by considering the contribution of the tension
term \cite{Tension,Kastor,Kurita,Traschen,Townsend}, then we
investigate the analogy of the laws of the AdS soliton solution. We
find the same analogy as that of the laws of the KK bubble. In
addition, we also investigate the laws of the boosted Ricci flat
black hole~\cite{Kastor2}. The boosted Ricci flat black hole can be
obtained from the static Ricci flat black hole by a boost
transformation along the compact spatial coordinate~\cite{Cai2}.
Note that, because the spatial coordinate is compact, the boosted
Ricci flat black hole is not equivalent to the static one
globally~\cite{Lemos,Awad,Cai2}. And these kind of globally
stationary but locally static spacetimes could be considered as the
gravitational analog of the Aharonow-Bohm
effect~\cite{Aharonow,Stachel}. Similarly, for the static AdS
soliton solution, we can also make a boost transformation along the
compact spatial coordinate of the static AdS soliton solution, and
then obtain the boosted AdS soliton solution. Like the AdS soliton
solution, the boosted AdS soliton solution also has the same
interchange symmetry with the above boosted Ricci flat black hole.
However, there are closed timelike curves and conical singularity in
the boosted AdS soliton solution, thus this solution is ill in
physics and the direct analogy of the energy and tension laws of the
boosted AdS soliton solution is of no sense. In spite of that, its
conserved charges such as the energy and momentum are well-defined,
and an interesting result is that the energy of the boosted AdS
soliton is lower than that of the AdS soliton. On the other hand,
note that although here we can easily find that the energy and
tension laws of the boosted Ricci flat black hole or the AdS soliton
solution are the same as those of the asymptotical flat case, the
underlying contents are not the same. First of all, the methods of
calculating the conserved charges are different. Because what they
discuss are the asymptotically flat cases, the well-known ADM
calculation can be used in their cases~\cite{Kastor,Kastor2}.
However, it is invalid and there have been several methods to
calculate the conserved charges in the asymptotically AdS
case~\cite{Counterterm,Euclidean,HH,AD,AM,Brown}. Here we just use
the surface counterterm method or Euclidean method. Second, they
obtain the laws by using the Hamiltonian perturbation theory
techniques~\cite{Traschen,Perturbation technique}, and more
expressive is that they should use the Hamiltonian formalism
presented by the ADM method~\cite{Kastor,Kastor2}. While for black
holes we obtain the laws just by applying the Euclidean
method~\cite{Euclidean}, and basing on this we obtain the laws of
AdS soliton by using the property of interchange symmetry. During
the derivation of laws, we do not need the explicit formalisms of
conserved charges. Thus, our results could also be considered as a
simple generalization of the results in asymptotically flat case to
the asymptotically AdS case~\cite{Kastor,Kastor2}.

The rest of paper is organized as follows. In section II, we
reinvestigate the energy and tension laws of the Ricci flat black
hole by considering the contribution of the tension term. In section
III, inspired from the interchange symmetry with the Ricci flat
black hole, we obtain the analogy of the laws of the AdS soliton
solution. In section IV, we generalize the above discussion in
section II to the case of the boosted Ricci flat black hole. In
section V, we consider the boosted AdS soliton solution. Finally, in
section VI, we give a brief conclusion and discussion.


\section{Reinvestigation of the energy and tension laws of the Ricci flat black hole%
} The so-called Ricci flat black hole solution considered here is
\cite{Horowitz,Maldacena}
\begin{equation}
ds^{2}=\frac{r^{2}}{l^{2}}[-(1-\frac{r_{0}^{4}}{r^{4}}%
)dt^{2}+dy^{2}+(dx^{i})^{2}]+(1-\frac{r_{0}^{4}}{r^{4}})^{-1}\frac{l^{2}}{%
r^{2}}dr^{2}.\ \ (i=1,2) \label{Ricci flat black hole}
\end{equation}%
which arises in the near-horizon geometry of p-brane and is
asymptotically the five dimensional AdS metric. It is easy to find
that its event horizon locates at $r_{+}=r_{0}$. And in order to
remove the conical singularity at the horizon, the Euclidean time
$\tau $ must have a period $\beta =\frac{\pi l^{2}}{r_{0}}$. Note
that, the coordinate $y$ is a compact spatial coordinate, and its
period is $\eta $. As the usual treatment, we can use the Euclidean
method to research the thermodynamics of the Ricci flat black
hole~\cite{Euclidean}. Choosing the pure AdS spacetime as the
reference background, we can easily obtain the Euclidean action of
the Ricci flat black hole
\begin{eqnarray}
I_{E}=-\frac{\beta r_{0}^{4}}{16\pi l^{5}}\eta V_{2}.
\label{Euclidean action1}
\end{eqnarray}
where $V_{2}$ is the coordinate volume of the surfaces parameterized by $%
x^{i} $. Thus, the free energy of the Ricci flat black hole
evaluated on the pure AdS background is~\cite{Euclidean}
\begin{eqnarray}
F &\equiv &\frac{I_{E}}{\beta }=E-TS=-\frac{ r_{0}^{4}}{16\pi
l^{5}}\eta V_{2}.  \label{free energy1}
\end{eqnarray}
And the energy and entropy are
\begin{eqnarray}
E &=&\frac{\partial I_{E}}{\partial \beta }=\frac{3r_{0}^{4}}{16\pi l^{5}}%
\eta V_{2}, \\
S &=&\beta \frac{\partial I_{E}}{\partial \beta
}-I_{E}=\frac{\eta V_{2}r_{0}^{3}}{4l^{3}}. \label{energyandentropy}
\end{eqnarray}
From~(\ref{energyandentropy}), it can be seen that the entropy $S$
is exactly equal to 1/4 of the horizon area $A$, which implicates
that those thermodynamical equations hold
\begin{eqnarray}
dF=-SdT, dE=TdS.  \label{masslaw1}
\end{eqnarray}
It should be emphasized that we have not considered the contribution
of tension term to the laws above, i.e gravitational tension. And it
is known that the gravitational tension term could contribute to the
first law in the case of the black p-branes or black string if the
size of the compact spatial coordinate is allowed to be changed. The
fact is that the geometry looks locally like the black string when
is far from the horizon of the Ricci flat black hole, thus the
gravitational tension term may also contribute to the
thermodynamical laws~\cite{Kurita}. And it is true that if assuming
the free energy in~(\ref{free energy1}) is also the function of
$\eta $, we can obtain not only the energy and entropy but also the
gravitational tension
\begin{eqnarray}
E &=&(\frac{\partial I_{E}}{\partial \beta })_{\eta }=\frac{3r_{0}^{4}}{%
16\pi l^{5}}\eta V_{2},  \notag \\
S &=&\beta (\frac{\partial I_{E}}{\partial \beta })_{\eta
}-I_{E}=\frac{\eta
V_{2}r_{0}^{3}}{4l^{3}},  \notag \\
\Gamma  &=&\frac{1}{\beta }(\frac{\partial I_{E}}{\partial \eta })_{\beta }=-%
\frac{r_{0}^{4}}{16\pi l^{5}}V_{2}.  \label{ESTension}
\end{eqnarray}
On the other hand, in a $d+1$ dimensional spacetime $\mathcal{M}$,
the conserved charge associated with the killing vector $\xi ^{\mu
}$ generating an isometry of the boundary geometry $\partial
\mathcal {M}$ defined through the quasilocal stress tensor
is~\cite{Counterterm,Brown}
\begin{equation}
Q_{\xi }=\int_{\Sigma }d^{d-1}x\sqrt{\sigma }(u^{\mu }T_{\mu \nu
}\xi ^{\nu }).   \label{Conservedcharge}
\end{equation}%
where $\Sigma $ is a spacelike hypersurface in the boundary
$\partial \mathcal {M}$, and $u^{\mu }$ is the timelike unit vector
normal to it. $\sigma _{ab}$ is the metric on $\Sigma $ defined as
\begin{equation}
\gamma _{\mu \nu }dx^{\mu }dx^{\nu }=-N_{\Sigma }^{2}dt^{2}+\sigma
_{ab}(dx^{a}+N_{\Sigma }^{a}dt)(dx^{b}+N_{\Sigma }^{b}dt)
\label{BoundaryADM}
\end{equation}
and $\gamma _{\mu \nu }$ is the metric on the boundary. Thus, the
energy related with the timelike killing vector $\xi ^{\mu }$ and
the momentum could be defined respectively as
\begin{eqnarray}
E &=&\int_{\Sigma }d^{d-1}x\sqrt{\sigma }N_{\Sigma }(u^{\mu }T_{\mu
\nu
}u^{\nu }), \label{Mass}
\\P_{a} &=&\int_{\Sigma }d^{d-1}x\sqrt{\sigma }\sigma _{ab}u_{\mu
}T^{b\mu }.  \label{Momentum}
\end{eqnarray}
According to the surface counterterm method, the quasilocal stress
tensor for the asymptotically $AdS_{5}$ solution
is~\cite{Counterterm}
\begin{equation}
T_{\mu \nu }=\frac{1}{8\pi }(\theta _{\mu \nu }-\theta \gamma _{\mu \nu }-%
\frac{3}{l}\gamma _{\mu \nu }-G_{\mu \nu }).  \label{stresstensor}
\end{equation}%
where all the above tensors refer to the boundary metric $\gamma
_{\mu \nu }$ defined on the hypersurface $r=constant$, and $G_{\mu
\nu }=R_{\mu \nu }-\frac{1}{2}R\gamma _{\mu \nu }$ is the Einstein
tensor of $\gamma _{\mu \nu }$, $\theta _{\mu \nu
}=-\frac{1}{2}(\nabla _{\mu }n_{\nu }+\nabla _{v}n_{\mu })$ is the
extrinsic curvature of the boundary with the normal vector $n^{\mu
}$ in the spacetime. Therefore, we can easily obtain the useful
quasi-local stress tensor of the Ricci flat black hole~(\ref{Ricci
flat black hole})
\begin{eqnarray}
8\pi T_{tt} &=&\frac{3r_{0}^{4}}{2l^{3}r^{2}}+...
\label{stresstensor1}
\end{eqnarray}%
And the energy is
\begin{eqnarray}
E =\frac{3r_{0}^{4}}{%
16\pi l^{5}}\eta V_{2}. \label{Energy1}
\end{eqnarray}
which is consistent with the above result in~(\ref{ESTension}). In
addition, the general definition of gravitational tension in a given
asymptotically translationally-invariant spatial direction (i.e.
$x$) of a $D$ dimensional space-time is~\cite{Tension}
\begin{equation}
\Gamma =\frac{1}{\Delta t}\frac{1}{8\pi }\int_{S_{x}^{\infty
}}[F(K^{(D-2)}-K_{0}^{(D-2)})-F^{\upsilon }p_{\mu \nu }r^{\nu }]
\label{TensionDefinition}
\end{equation}%
here $S_{x}^{\infty }=\Sigma _{x}\cap \Sigma ^{\infty }$ and $\Sigma
_{x}$ is the hypersurface $x=const$ with unit normal vector $n^{\mu
}$, and $\Sigma ^{\infty }$ is the asymptotic boundary of the
spacetime with unit normal vector $r^{\mu }$. The space-like killing
vector $X^{\mu }$ corresponding to the translationally-invariant
spatial direction $x$ is decomposed into normal and tangential parts
to $\Sigma _{x}$ that
\begin{equation}
X^{\mu }=Fn^{\mu }+F^{\mu \label{Decomposition}}
\end{equation}
and the extrinsic curvature tensor on $\Sigma _{x}$ with respect to $n^{\mu }$ is $%
K_{\mu \nu }$, while $K^{(D-2)}$ is the extrinsic curvature of the surface $%
S_{x}^{\infty }$ in $\Sigma _{x}$, and $K_{0}^{(D-2)}$ is the
corresponding
extrinsic curvature of the surface $S_{x}^{\infty }$ in the reference space $%
(M,(g_{0})_{\mu \nu })$. The metric with respect to $n^{\mu }$ on
$\Sigma _{x}$ is
\begin{equation}
h_{\mu \nu }=g_{\mu \nu }-n_{\mu }n_{\nu \label{ReducedMetric}}
\end{equation}%
while the corresponding canonical momentum $p_{\mu \nu }$ with
respect to $h_{\mu \nu }$ is
\begin{equation}
p^{\mu \nu }=\frac{1}{\sqrt{h}}\pi ^{\mu \nu }=K^{\mu \nu }-Kh^{\mu
\nu \label{CanonicalMomentum}}
\end{equation}
Thus, from this general definition of gravitational
tension~(\ref{TensionDefinition}), we can obtain the gravitational
tension along the compact spatial direction $y$ in Ricci flat black
hole~(\ref{Ricci flat black hole})
\begin{eqnarray}
\Gamma  =-\frac{r_{0}^{4}}{16\pi l^{5}}V_{2}. \label{Tension1}
\end{eqnarray}
which is also consistent with the above result in~(\ref{ESTension}).
And these consistences of energy, tension and entropy implicate that
after adding the contribution of tension term the first laws
in~(\ref{masslaw1}) are
\begin{eqnarray}
dF &=&-SdT+\Gamma d\eta =-\frac{1}{8\pi }A_{H}d\kappa _{H}+\Gamma
d\eta ,
\notag \\
dE &=&TdS+\Gamma d\eta =\frac{1}{8\pi }\kappa _{H}dA_{H}+\Gamma
d\eta . \label{Masslaw2}
\end{eqnarray}
where $T=1/\beta =\kappa _{H}/2\pi $ and $S=A_{H}/4$. Using these
conserved charges, we can also easily check that
\begin{equation}
E-TS=\Gamma \eta.   \label{Smarr}
\end{equation}%
which is very similar with the Smarr relation. Thus
from~(\ref{Smarr}) and ~(\ref{Masslaw2}), we can obtain the tension
law that
\begin{equation}
\eta d\Gamma =-SdT.  \label{Tensionlaw1}
\end{equation}%
which can be found to have the same formalism with the static
Kaluza-Klein black hole which is asymptotically flat in
Refs~\cite{Kastor,Kastor2}.

\section{The AdS soliton solution, interchange symmetry, and analogy of
energy and tension laws}

The AdS soliton solution is~\cite{Horowitz}
\begin{equation}
ds^{2}=\frac{r^{2}}{l^{2}}[(1-\frac{r_{0}^{4}}{r^{4}}%
)dy^{2}-dt^{2}+(dx^{i})^{2}]+(1-\frac{r_{0}^{4}}{r^{4}})^{-1}\frac{l^{2}}{%
r^{2}}dr^{2}.\ \ (i=1,2)  \label{AdSSoliton}
\end{equation}%
with the coordinate $r$ restricted to $r\geq r_{0}$. Again, the
coordinate $y $ could be identified with period $\eta =\frac{\pi
l^{2}}{r_{0}}$ to avoid a conical singularity at $r=r_{0}$. Note
that this spacetime is completely nonsingular and globally static.
And it can be obtained from the Ricci flat black hole
metric~(\ref{Ricci flat black hole}) with the double analytic
continuation such that
\begin{equation}
t\rightarrow iy,y\rightarrow it.  \label{Double analytic
continuation}
\end{equation}%
which arises an interesting interchange symmetry between the AdS soliton
solution and the Ricci flat black hole.

Using the same surface counterterm method, we can calculate the
useful quasilocal stress tension of AdS soliton~\cite{Counterterm}
\begin{eqnarray}
8\pi T_{tt} &=&-\frac{r_{0}^{4}}{2l^{3}r^{2}}+...
\label{Stresstension2}
\end{eqnarray}%
Thus, the energy is
\begin{equation}
E=-\frac{r_{0}^{4}}{16\pi l^{5}}\eta V_{2}.  \label{Energy2}
\end{equation}%
In addition, the tension of the AdS soliton (along the direction of
the compact coordinate $y$) from the general
definition~(\ref{TensionDefinition}) is
\begin{equation}
\Gamma =\frac{3r_{0}^{4}}{16\pi l^{5}%
} V_{2}.  \label{Tension2}
\end{equation}%
Eqs~(\ref{Energy2})~(\ref{Tension2}) can explicitly manifest the
interchange symmetry with the Ricci flat black hole compared with
its energy and tension.

In the above section, during deriving the laws of Ricci flat black
hole, we mainly base on an underlying assumption that the
equations~(\ref{Masslaw2}) hold, and then find the consistence with
the calculations by other methods. However, for the AdS soliton
solution, the first problem is these equations may not hold because
the period of the imaginary time which usually relates with the
temperature is arbitrary in the AdS soliton spacetime. Moreover, if
we take the entropy just as the usual Bekenstein-Hawking entropy (it
is the 1/4 of the horizon area), we could find the entropy is zero.
Thus, the direct analogy of the mass and tension laws of the AdS
soliton solution like~(\ref{Masslaw2})~(\ref{Tensionlaw1}) seems to
be absent. In spite of that, inspired from the interchange symmetry
between the black hole and AdS soliton, it may have the analogy. And
it is true that it has been found the similar analogy of the KK
bubble in Ref~\cite{Kastor} where it discusses the asymptotically
flat case. As same as that of KK bubble, we can also first define
some new quantities, such as the surface gravity and the area of the
AdS soliton. And according to these definitions, the surface gravity
and the area of the AdS soliton are~\cite{Kastor}
\begin{equation}
\kappa _{s}=\frac{2r_{0}}{l^{2}},A_{s}=\frac{V_{2}r_{0}^{3}}{l^{3}}.
\label{SurfacegravityandArea}
\end{equation}%
However, here we would not use the Hamiltonian perturbation
techniques to deduce the laws of AdS soliton until one finds its
appropriate formalisms of the conserved charges and gravitational
tension as those of KK bubble. And we just base on its interchange
symmetry with the Ricci flat black hole~(\ref{Ricci flat black
hole}). From the quantities in~(\ref{SurfacegravityandArea}) and
those in ~(\ref{Energy2})~(\ref{Tension2}), we can make an easy
displacement in~(\ref{Masslaw2})~(\ref{Smarr}) and
~(\ref{Tensionlaw1}) by using the interchange symmetry such that
\begin{equation}
E\rightarrow \Gamma \eta ,T\rightarrow T,S\rightarrow S\eta ,\Gamma
\rightarrow E/\eta.  \label{displacement}
\end{equation}
Thus we can obtain the reduced relations
\begin{equation}
d\Gamma =\frac{1}{8\pi G}\kappa _{s}dA_{s}.  \label{MassTesionlaw3}
\end{equation}
\begin{equation}
dE=-\frac{1}{8\pi G}\eta A_{s}d\kappa _{s}+(\Gamma -\frac{1}{8\pi
G}\kappa _{s}A_{s})d\eta .  \label{12}
\end{equation}%
From which, we can easily check out that they hold by using the
quantities
in~(\ref{Energy2})~(\ref{Tension2})~(\ref{SurfacegravityandArea})
and see that they have the similar formalisms with the laws of black
hole. Thus they can be naturally considered as the analogy of the
energy and tension laws of the AdS soliton. The most interesting
thing is that they has the same formalism with the result of the K-K
bubble in Ref~\cite{Kastor} where it is deduced by using the
Hamiltonian perturbation theory
techniques~\cite{Traschen,Perturbation technique}. Thus, it is more
convincible that they could be considered as the analogy of the
energy and tension laws of the AdS soliton.

\section{The energy and tension laws for boosted Ricci flat black hole}

The boosted Ricci flat black hole can be obtained from~(\ref{Ricci
flat black hole}) by the following boost transformation~\cite{Cai2}
\begin{eqnarray}
t &\rightarrow &t\cosh \alpha -y\sinh \alpha ,  \nonumber \\
y &\rightarrow &-t\sinh \alpha +y\cosh \alpha .  \label{Boost
transformation}
\end{eqnarray}%
where $\alpha $ is the boost parameter and the boost velocity is
$v=\tanh \alpha $. Thus, the metric of the boosted Ricci flat black
hole is
\begin{equation}
ds^{2}=\frac{r^{2}}{l^{2}}[-dt^{2}+dy^{2}+\frac{r_{0}^{4}}{r^{4}}(dt\cosh
\alpha -dy\sinh \alpha )^{2}+(dx^{i})^{2}]+(1-\frac{r_{0}^{4}}{r^{4}})^{-1}%
\frac{l^{2}}{r^{2}}dr^{2}.\ \ (i=1,2)  \label{Boosted Ricci flat
black hole}
\end{equation}%
Note that, because the coordinate $y$ in~(\ref{Boost
transformation}) is periodic, the solution~(\ref{Boosted Ricci flat
black hole}) is not equivalent to the static Ricci flat black
hole~(\ref{Ricci flat black hole}) globally~\cite{Lemos,Awad,Cai2}.
And in order to remove the conical singularity at the horizon
$r=r_{0}$, the Euclidean time $\tau $ in~(\ref{Boosted Ricci flat
black hole}) could have a period $\beta =\frac{\pi l^{2}\cosh \alpha
}{r_{0}}$. Following the same procedure as section II, at first we
do not consider the contribution from the gravitational tension term
in the laws of thermodynamics. After choosing the pure AdS spacetime
as the background and using the same Euclidean method, we can obtain
the Euclidean action of the boosted Ricci black hole to
be~\cite{Euclidean}
\begin{equation}
I_{E}=-\frac{\beta r_{0}^{4}}{16\pi Gl^{5}}\eta V_{2}.
\label{Euclidean action2}
\end{equation}
Note that, although the Euclidean action is the same as that of the
static Ricci flat black hole~(\ref{Euclidean action1}), the
relationship between $\beta$ and $r_{0}$ is different. Moreover,
here the thermal function related with Euclidean action is the
Gibbons free energy~\cite{Euclidean}
\begin{eqnarray}
G &\equiv &\frac{I_{E}}{\beta }=E-TS-vP.  \label{Gibbons Free
energy}
\end{eqnarray}
and $G$ is the function of not only $\beta$ but also the boost
velocity $v$.  Thus, the energy, entropy and momentum are
\begin{eqnarray}
E &=&(\frac{\partial I_{E}}{\partial \beta })_{v}-\frac{v}{\beta }(\frac{%
\partial I_{E}}{\partial v})_{\beta }=\frac{(3+4a^{2})r_{0}^{4}}{16\pi l^{5}}%
\eta V_{2},  \notag \\
S &=&\beta (\frac{\partial I_{E}}{\partial \beta
})_{v}-I_{E}=\frac{\eta
V_{2}r_{0}^{3}}{4l^{3}}\sqrt{1+a^{2}},  \notag \\
P &=&-\frac{1}{\beta }(\frac{\partial I_{E}}{\partial v})_{\beta }=\frac{%
\eta V_{2}r_{0}^{4}}{4\pi l^{5}}a\sqrt{1+a^{2}}.
\label{EnergyEntropyMomentum}
\end{eqnarray}
where $a\equiv \sinh \alpha $ and it could be easily seen that the
entropy $S$ is also exactly equal to 1/4 of the horizon area $A$,
which implicates that the following relations hold
\begin{eqnarray}
dG &=&-SdT-Pdv,  \notag \\
dE &=&TdS+vdP.  \label{Masslaw3}
\end{eqnarray}
Again if assuming the Gibbons free energy $G$ in~(\ref{Gibbons Free
energy}) is also the function of $\eta $, we can also obtain the
gravitational tension
\begin{eqnarray}
E &=&(\frac{\partial I_{E}}{\partial \beta })_{v,\eta }-\frac{v}{\beta }(%
\frac{\partial I_{E}}{\partial v})_{\beta ,\eta }=\frac{(3+4a^{2})r_{0}^{4}}{%
16\pi l^{5}}\eta V_{2},  \notag \\
S &=&\beta (\frac{\partial I_{E}}{\partial \beta })_{v,\eta }-I_{E}=\frac{%
\eta V_{2}r_{0}^{3}}{4l^{3}}\sqrt{1+a^{2}},  \notag \\
P &=&-\frac{1}{\beta }(\frac{\partial I_{E}}{\partial v})_{\beta ,\eta }=%
\frac{\eta V_{2}r_{0}^{4}}{4\pi l^{5}}a\sqrt{1+a^{2}},  \notag \\
\Gamma  &=&\frac{1}{\beta }(\frac{\partial I_{E}}{\partial \eta
})_{\beta ,v}=-\frac{r_{0}^{4}}{16\pi l^{5}}V_{2}. \label{ESP2}
\end{eqnarray}
On the other hand, according to the definition, the useful
quasi-local stress tensor of the boosted Ricci flat black
hole~(\ref{Boosted Ricci flat black hole}) is~\cite{Counterterm}
\begin{eqnarray}
8\pi T_{tt} &=&\frac{(3+4\sinh ^{2}\alpha
)r_{0}^{4}}{2l^{3}r^{2}}+...
\nonumber \\
8\pi T_{ty} &=&-\frac{2\sinh \alpha \cosh \alpha
r_{0}^{4}}{l^{3}r^{2}}+... \label{StressTensor2}
\end{eqnarray}%
From which the energy and momentum can be calculated to be
\begin{eqnarray}
E &=&\frac{(3+4\sinh ^{2}\alpha )r_{0}^{4}}{16\pi l^{5}}\eta V_{2},
\nonumber \\
P &=&\frac{\sinh \alpha \cosh \alpha r_{0}^{4}}{4\pi l^{5}}\eta
V_{2}. \label{MPTension}
\end{eqnarray}
where the energy and momentum are consistent with the above results
in~(\ref{ESP2}). And this consistence could implicate that after
adding the contribution of tension term the first laws
in~(\ref{Masslaw3}) are
\begin{eqnarray}
dG &=&-SdT-Pdv+\Gamma d\eta ,  \notag \\
dE &=&TdS+vdP+\Gamma d\eta .  \label{Masslaw4}
\end{eqnarray}
However, if we use the general definition of gravitational
tension~(\ref{TensionDefinition}), we can obtain the tension
\begin{equation}
\Gamma^{'} =-\frac{(1+4\sinh ^{2}\alpha )r_{0}^{4}}{16\pi
l^{5}}V_{2}. \label{Tension3}
\end{equation}
which is not consistent with the result in~(\ref{ESP2}). Note that,
this difference has also been found by D. Kastor et al, and they
argued that the tension obtained in~(\ref{ESP2}) was in fact an
effective tension which was related to the general tension such
that~\cite{Kastor2}
\begin{equation}
\Gamma =\Gamma^{'} +\frac{vP}{\eta }.  \label{TTRelation}
\end{equation}%
From which, we can also find that when the boosted velocity is zero,
the general tension is just equal to the effective tension.

Using these quantities in ~(\ref{ESP2})~(\ref{MPTension}), we can
also check that
\begin{equation}
E-TS-vP=\Gamma \eta   \label{Smarr2}
\end{equation}
Thus, from this relation~(\ref{Smarr2}) and the first energy
law~(\ref{Masslaw4}), the first tension law of boosted Ricci flat
black hole is
\begin{equation}
SdT+Pdv+\eta d\Gamma =0  \label{Tensionlaw2}
\end{equation}

\section{The boosted AdS soliton solution}
Naturally, we can also make a boost transformation~(\ref{Boost
transformation}) along the compact coordinate $y$ in the static AdS
soliton solution~(\ref{AdSSoliton}). Thus, the boosted AdS soliton
solution is
\begin{equation}
ds^{2}=\frac{r^{2}}{l^{2}}[-dt^{2}+dy^{2}-\frac{r_{0}^{4}}{r^{4}}(dy\cosh
\alpha -dt\sinh \alpha )^{2}+(dx^{i})^{2}]+(1-\frac{r_{0}^{4}}{r^{4}})^{-1}%
\frac{l^{2}}{r^{2}}dr^{2}.\ \ (i=1,2)  \label{BoostedAdSSoliton}
\end{equation}
Note that this solution is also different from the static AdS
soliton globally, and it is easy to see that the coordinate $y$ in
the boost transformation is compact. In addition, an interesting
result is that this boosted AdS soliton solution also has the same
interchange symmetry with the boosted Ricci flat black hole. That
is, it can also be obtained from the boosted Ricci flat black hole
solution by the double analytic continuation between the time and
the compact coordinate $y$ in~(\ref{Boosted Ricci flat black hole}).
In the above section we have obtained the analogy of the energy and
tension laws of the static AdS soliton solution through the
inspiration from the interchange symmetry with the Ricci flat black
hole. However, it's easily found that there are closed timelike
curves in the boosted AdS soliton
solution~(\ref{BoostedAdSSoliton}). Moreover, viewed from the
physical point, after boosting along the compact coordinate $y$ in
static AdS soliton~(\ref{AdSSoliton}), the period of $y$ would be
shrunk to $\gamma ^{-1}\eta $ where $\gamma =(1-v^{2})^{-1/2}=\cosh
\alpha $ is the shrinking factor. However, the new period could not
avoid the conical singularity. Thus, this boosted AdS soliton
solution is ill in physics and the direct analogy of laws is of no
sense. In spite of that, the conserved charges such as energy and
momentum are well defined because they just depend on the properties
of its asymptotic behavior. And the corresponding quasi-local stress
tensor of the boosted AdS soliton solution can be
obtained~\cite{Counterterm}
\begin{eqnarray}
8\pi T_{tt} &=&-\frac{(1+4\sinh ^{2}\alpha
)r_{0}^{4}}{2l^{3}r^{2}}+...
\nonumber \\
8\pi T_{ty} &=&\frac{2\sinh \alpha \cosh \alpha
r_{0}^{4}}{l^{3}r^{2}}+... \label{StressTensor4}
\end{eqnarray}%
Thus, the energy and momentum of the boosted AdS soliton solution
are
\begin{eqnarray}
E &=&-\frac{(1+4\sinh ^{2}\alpha )r_{0}^{4}}{16\pi l^{5}}\eta V_{2},
\nonumber \\
P &=&-\frac{\sinh \alpha \cosh \alpha r_{0}^{4}}{4\pi l^{5}}\eta
V_{2}. \label{EP4}
\end{eqnarray}%
In addition, the general tension can also be obtained from the
definition~(\ref{TensionDefinition})
\begin{equation}
\Gamma =\frac{(3+4\sinh ^{2}\alpha )r_{0}^{4}}{16\pi l^{5}}V_{2}.
\label{T4}
\end{equation}%
These quantities in~(\ref{EP4})~(\ref{T4}) could explicitly manifest
the interchange symmetry with the boosted Ricci flat black hole,
too.

\section{Conclusion and discussion}

One of the motivations of this paper is to obtain the analogy of the
energy and tension laws of the AdS soliton solution, which can give
more understanding of this solution. In order to obtain them, we
first reconsider the laws of the Ricci flat black hole by taking the
contribution of the tension term into account. Then, inspired from
the interchange symmetry between the Ricci flat black hole and AdS
soliton, we finally obtain the analogy. In spite of that, how to
understand the analogy of laws of the AdS soliton is an open
question. Particularly, whether there is some underlying physical
interpretations such as thermodynamical effects in it is worthy of
further discussion. In addition, as a more general asymptotically
AdS black hole solution, we also take the boosted Ricci flat black
hole for example to give a simple generalization of the works by
D.Kastor to the asymptotically AdS case. Note that, although here
our formalisms of the laws of black holes or the static soliton are
the same as those of the asymptotically flat cases, the underlying
deduced contents are different. In principle, if we find the
appropriate formalisms of conserved charges and gravitational
tension, we perhaps can also use the Hamiltonian perturbation method
to deduce these laws directly. And this possibility will be
considered in the future work. As the corresponding solution which
has the interchange symmetry with boosted Ricci flat black hole, we
also consider the boosted AdS soliton solution. However, although
there is the same interchange symmetry, this boosted AdS soliton
solution is ill in physics because of the existence of the closed
timelike curves and conical singularity. Thus, the direct analogy of
energy and tension laws are of no sense. In spite of that, an
interesting result is that the conserved charges such as the energy
and momentum are well-defined for the boosted AdS soliton solution.
Moreover, as we expected, its energy is smaller than that of the
static AdS soliton solution. Thus, whether it can be considered as a
violation case to the new positive energy conjecture proposed by G.T
Horowitz and R.C Myers and how to understand it from the viewpoint
of the AdS/CFT correspondence would also be interesting things to
give further discussions. In addition, during calculating the
conserved charges, we also find that perhaps there is a new way to
define the gravitational tension from the quasi-local stress tensor
defined in~(\ref{stresstensor}), because the gravitational tension
can be easily found to be related to the corresponding stress tensor
$T_{yy}$ such that
\begin{eqnarray}
\text{Ricci flat black hole} &\text{: }&\Gamma =-\frac{r_{0}^{4}}{16\pi l^{5}%
}V_{2},\text{ }T_{yy}=\frac{r_{0}^{4}}{16\pi l^{3}r^{2}}+....\text{\
}
\notag \\
\text{Static AdS soliton} &\text{:}&\Gamma =\frac{3r_{0}^{4}}{16\pi l^{5}}%
V_{2},\text{ }T_{yy}=-\frac{3r_{0}^{4}}{16\pi
l^{3}r^{2}}+....\text{\ }
\notag \\
\text{Boosted Ricci flat black hole} &\text{:}&\Gamma =-\frac{%
(1+4a^{2})r_{0}^{4}}{16\pi l^{5}}V_{2},\text{ }T_{yy}=\frac{%
(1+4a^{2})r_{0}^{4}}{16\pi l^{3}r^{2}}+....  \notag \\
\text{Boosted AdS soliton} &\text{:}&\Gamma =\frac{3r_{0}^{4}}{16\pi l^{5}}%
V_{2},\text{ }T_{yy}=-\frac{(3+4a^{2})r_{0}^{4}}{16\pi
l^{3}r^{2}}+.... \label{TTYYRelation}
\end{eqnarray}
On the other hand, viewed from the physical interpretation of the
stress tensor, its spatial diagonal components are related with the
pressure, thus it is more convincible that there is a new
possibility to define the gravitational tension. In fact,
considering the interchange symmetry and the formalisms
in~(\ref{Mass})~(\ref{TensionDefinition}), we can give the new
definition of the gravitational tension through the counterterm in
the asymptotical AdS case that
\begin{equation}
\Gamma =-\frac{1}{\Delta t}\int_{S_{x}^{\infty }}d^{d-1}x\sqrt{\sigma }%
F(n^{\mu }T_{\mu \nu }n^{\nu })  \label{TensionNewDefinition}
\end{equation}
which can be easily checked that this new definition is satisfied in
our cases.

Note that, after our paper appeared, Dr. Cristian Stelea showed me
that they had also already given an exact formalism of the
gravitational tension through the counterterm in their cases. Thus
giving a more general rigorous definition of the gravitational
tension through the counterterm is an open interesting question, and
perhaps some clues could be found in their works~\cite{Stelea}.

\section{Acknowledgements}

Y.P Hu thanks Professor Rong-Gen Cai and Dr.Li-Ming Cao, Jia-Rui
Sun, Xue-Fei Gong and Chang-Yong Liu for their helpful discussions.
And Y.P Hu also thanks Dr. Cristian Stelea for his useful
information. This work is supported partially by grants from NSFC,
China (No. 10325525, No. 90403029 and No. 10773002), and a grant
from the Chinese Academy of Sciences.

\end{document}